\documentclass[12pt,a4paper]{article}

\usepackage{latexsym,graphicx}

\title{Extreme objects with arbitrary large mass,\\
 or density, and arbitrary size}

\author{J.\ M.\ Aguirregabiria and Ll.\ Bel\\
\emph{Fisika Teorikoa, Euskal Herriko Unibertsitatea},
\\\emph{P.K. 644, 48080 Bilbo, Spain}
}

\begin{document}
\maketitle

\begin{abstract}

We consider a generalization of the interior Schwarzschild  solution that
we match to the exterior one to build global $\mathcal{C}^1$ models that
can have arbitrary large mass, or density, with arbitrary size.
This is possible because of a new insight into the problem of localizing
the center of symmetry of the models and the use of principal
transformations to understand the structure of space.

\end{abstract}


\section{Introduction}
\label{sec:intro}

We consider in this paper a family of spherically symmetric, static
models with bounded sources. To start with we shall consider the
following reduced form of the line-element:
\begin{equation}\label{1.1}
ds^2= -A^2dt^2+d\hat s^2,
\end{equation}
with
\begin{equation}\label{1.2}
d\hat s^2=B^2dr^2+BCr^2d\Omega^2, \qquad
d\Omega^2=d\theta^2+\sin^2\theta d\varphi^2,
\end{equation}
$A$, $B$, and $C$ being three functions of $r$. This form of the
line-element is fully adapted to the assumption requiring the
existence of a global time-like integrable Killing vector field, as well as to
the assumption of spherical symmetry. It remains invariant under  the
adapted time coordinate transformation
\begin{equation}\label{1.3}
\bar t=Kt+K_0,
\end{equation}
where $K$ is an arbitrary positive constant which is usually chosen
such that
\begin{equation}\label{1.4}
\lim_{r\to\infty}A=1,
\end{equation}
and $K_0$ is a second arbitrary constant that can be completely
ignored. It is also invariant under a radial coordinate transformation

\begin{equation}\label{1.5}
\bar r= \bar r(r),
\end{equation}
which may serve different purposes: mathematical simplicity or desired
interpretation.

Among the coordinate conditions that can be used to choose
$r$ one finds:

i) The historical condition
\begin{equation}\label{1.0}
A^2B^4C^2r^4=1.
\end{equation}
This is equivalent to the coordinate condition used by Schwarzschild to
derive his exterior solution.

ii) The curvature condition
\begin{equation}\label{1.6}
BC=1.
\end{equation}
It was first considered for mathematical simplicity by Droste
\cite{Droste}, Hilbert~\cite{Hilbert} and Weyl~\cite{Weyl} and it is the
most used.

iii) The isotropic condition
\begin{equation}\label{1.7}
C=B.
\end{equation}

iv) The harmonic condition
\begin{equation}\label{1.8}
B=\frac{1}{2rA}\left(r^2AC\right)',
\end{equation}
where the prime means a derivative with respect to $r$. This condition
is equivalent to requiring that the three coordinates
\begin{equation}\label{1.9}
x^1=r\sin\theta\cos\varphi, \qquad x^2=r\sin\theta\sin\varphi, \qquad
x^3 =r\cos\theta
\end{equation}
be harmonic functions in the space-time defined by~(\ref{1.1}), i.e.,
\begin{equation}\label{1.10}
\triangle x^k=
\frac{1}{\sqrt{-g}}\partial_i\left(\sqrt{-g}g^{ij}\partial_jx^k\right)=0, \qquad
g=\det\left(g_{\alpha\beta}\right).
\end{equation}
(Greek indices run from 0 to 3 and Latin indices from 1 to 3.)
This condition which leaves open the choice of an arbitrary constant
was used, with a particular value of this constant, by Fock~\cite{Fock} who meant
to give a particular meaning to this particular radial coordinate $r$.

v) The Gauss condition:
\begin{equation}\label{1.10a}
B=1.
\end{equation}

In recent years one of us (Ll. B) has introduced the quo-harmonic
condition~\cite{Bel95a}:
\begin{equation}\label{1.11}
B=\frac{1}{2r}\left(r^2C\right)^\prime.
\end{equation}
This is equivalent to requiring that the three functions~(\ref{1.9}) be
harmonic in the 3-dimensional space with line-element~(\ref{1.2}), i.e.,
\begin{equation}\label{1.12}
\hat\triangle x^k=
\frac{1}{\sqrt{\hat g}}\partial_i\left(\sqrt{\hat g}\hat
g^{ij}\partial_jx^k\right)=0, \qquad \hat g=\det\left(g_{ij}\right), \qquad
\hat g^{ij}=g^{ij}.
\end{equation}

This new coordinate condition has been thoroughly discussed in
\cite{Bel95b,ABMMR}. It is intimately connected  with the concept
of principal transformation of a 3-dimensional Riemannian metric
\cite{Bel96}.
Although not every such metric possesses a principal transform, many
do have it and in particular those with spherical symmetry. In this
particular case a principal transform of~(\ref{1.2}) is a new metric
with line-element
\begin{equation}\label{1.13}
d\bar s^2=\Phi^2B^2dr^2+\Psi^2BCr^2d\Omega^2,
\end{equation}
where $\Phi$ and $\Psi$ are two functions of $r$ such that

i) the Riemann tensor of~(\ref{1.13}) is zero,
\begin{equation}\label{1.14}
\bar R_{ijkl}=0,
\end{equation}
i.e., the metric is flat, and

ii) the transformation from~(\ref{1.2}) to~(\ref{1.13}) is harmonic:
\begin{equation}\label{1.15}
\left(\hat\Gamma^i_{jk}-\bar\Gamma^i_{jk}\right)\hat g^{jk}=0,
\end{equation}
where $\hat\Gamma^i_{jk}$ and $\bar\Gamma^i_{jk}$ are respectively
the second-kind Christoffel symbols of~(\ref{1.2}) and~(\ref{1.13}). Since
both~(\ref{1.14}) and~(\ref{1.15}) are tensor equations as long as only
coordinates that are adapted to the main Killing vector are
considered, both conditions are intrinsic to the static character of
the models.

To solve for $\Phi$ and $\Psi$ two methods can be used. The first one
consist in using directly Eqs.~(\ref{1.14}) and~(\ref{1.15}) where these
two functions are the unknowns. It has the advantage of allowing to
use any coordinate condition that one wishes. The second method is
based on the remark that if we consider the problem solved and we use
Cartesian coordinates $x^i$ of~(\ref{1.13}), i.e., such that
\begin{equation}\label{1.16}
\bar\Gamma^i_{jk}=0,
\end{equation}
then from~(\ref{1.15}) it follows that in this system of coordinates  we shall have
\begin{equation}\label{1.17}
\hat\Gamma^i_{jk}g^{jk}=0.
\end{equation}
These conditions tell us that $x^i$ is a system of
quo-harmonic coordinates of~(\ref{1.2}), i.e., solutions of~(\ref{1.12}).
>From this remark it follows that if we solve first these equations
for the functions
\begin{equation}\label{1.18}
x^1=R(r)\sin\theta\cos\varphi, \qquad x^2=R(r)\sin\theta\sin\varphi, \qquad
x^3 =R(r)\cos\theta,
\end{equation}
that is to say, if we first solve the single equation for the unknown
function~$R(r)$
\begin{equation}\label{1.19}
r^2R''+r\left(2-rB^{-1}B'\right)R'-2B^2R=0,
\end{equation}
then to solve Eqs.~(\ref{1.14}) and~(\ref{1.15}) we just have to equate the
line-element of flat space written in polar coordinates $R, \theta,
\varphi$ with~(\ref{1.13}):
\begin{equation}\label{1.20}
dR^2+R^2d\Omega^2=\Phi^2B^2dr^2+\Psi^2BCr^2d\Omega^2,
\end{equation}
whence it follows that
\begin{equation}\label{1.21}
\Phi=\frac{R'}{B}, \qquad \Psi=\frac{R}{r\sqrt{BC}}.
\end{equation}

Principal transformations were introduced in~\cite{Bel96} as a
generalisation to 3-dimensional Riemannian metrics of one of Gauss's
theorems according to which any 2-dimensional metric can be mapped
conformally, locally, into an Euclidean space. In this paper they
will play also an important role in the interpretation of the models
to be presented in Sect.~\ref{sec:model}.

These models will be required to satisfy the following conditions:

i) A value of the coordinate $r$ exists, say $r_1$, such that for
$r>r_1$ the space-time model is a solution of the vacuum field
equations
\begin{equation}\label{1.22}
S_{\alpha\beta}=0,
\end{equation}
where $S_{\alpha\beta}$ is the Einstein tensor. It is therefore
the exterior Schwarzschild solution.

ii) For $r<r_1$ the space-time model is an interior solution with a
perfect-fluid source
\begin{equation}\label{1.23}
S_{\alpha\beta}=\kappa T_{\alpha\beta}, \qquad \kappa=8\pi,\qquad
T_{\alpha\beta}=(\rho+p)u_\alpha u_\beta+pg_{\alpha\beta},
\end{equation}
where $\rho$ is the energy density, $p$ the pressure and $u^\alpha$ the
4-velocity of the fluid, tangent to the main Killing field. The solution
that we shall consider is a rather straightforward, but physically
innovating, generalization of the interior Schwarzschild solution. This
generalization is crucial to the goal that we pursue in this paper.
Namely the possibility of constructing global models with arbitrary
mass or arbitrary density, and arbitrary size.

iii) On the border $r=r_1$ we shall require the continuity of the
three functions $A, B, C$ as well as the continuity of the three
derivatives $A',B',C'$ thus completing the
construction of models of class $\mathcal{C}^1$. This requirement greatly
restricts the choice of a unique global coordinate condition, i.e.,
being the same on both sides of the border.


\section{The exterior Schwarzschild solution}
\label{sec:exterior}

The exterior Schwarzschild solution has been described using
the coordinate conditions i) to iv) mentioned in the preceding
section. We list below the corresponding expressions of the
coefficients $A$, $B$ and $C$, and the intervals of the corresponding
radial coordinate $r$ on which the metric is static.

i) Historical condition:%
\footnote{Schwarzschild's original solution has been reconsidered
recently in Ref.~\cite{Antoci}.}
\begin{equation}\label{2.1}
A=\sqrt{1-\frac{2m}{R}}, \qquad B=A^{-1}R', \qquad \sqrt{BC}=R/r,
\qquad r\in \left] 0,\infty\right[,
\end{equation}
where $R(r)$ is the function
\begin{equation}\label{2.2}
R(r)=\left[r^3+(2m)^3\right]^{1/3}.
\end{equation}

ii) Curvature condition:
\begin{equation}\label{2.3}
A=\sqrt{1-\frac{2m}{r}}, \qquad B=A^{-1}, \qquad BC=1,
\qquad r\in \left]2m,\infty\right[.
\end{equation}

iii) Isotropic condition:
\begin{equation}\label{2.4}
A=\frac{1-m/{2r}}{1+m/{2r}}, \qquad
B=C=\left(1+\frac{m}{2r}\right)^2, \qquad r\in \left]m/2,\infty\right[.
\end{equation}

iv) Particular harmonic condition:

\begin{equation}\label{2.5}
A=\sqrt{\frac{r-m}{r+m}}, \qquad B=A^{-1}, \qquad BC=\left(1+\frac{m}{r}\right)^2,
\qquad r\in \left]m,\infty\right[.
\end{equation}

v) Gauss condition:

\begin{equation}\label{eq:gauss}
A=\sqrt{1-\frac{2m}{R}}, \qquad B=1, \qquad C=R,
\qquad r\in \left]D,\infty\right[,
\end{equation}
where $D$ is the arbitrary constant  in the following definition of $R$:
\begin{equation}
R\equiv r \sqrt{1-\frac{2m}{r}}+m \ln\left(-1+\frac{r}{m}+\frac{r}{m}\sqrt{1-\frac{2m}{r}}\right)+D.
\end{equation}

The lowest value of $r$, say $r_B$, in the domain of staticity, different for
each of the coordinate conditions considered so far, defines a
border, say $\mathcal{B}$,
which was used to be called the Schwarzschild singularity and that
now is more often called the horizon of the exterior Schwarzschild 
solution.

In the case of the original Schwarzschild coordinate, but also
when using the Gaussian condition and setting $D=0$, 
one has $r_B=0$, strongly suggesting that Schwarzschild had succeeded, as it was his
intention, in identifying the most extreme source of his solution as being
a point. This point of view it is still accepted in
\cite{Abrams} where
the author claims that the non zero values of
$r_B$, suggesting that $\mathcal{B}$ is in fact a 2-dimensional surface,
come from an inadequate choice of the coordinate condition.
 We want to
comment that whether $\mathcal{B}$ is a point or a surface depends
intrinsically on the geometry which is describing the structure of
the space in the static reference frame. If the metric of this space is
that with line-element~(\ref{1.2}),
i.e., if one accepts that the physical distance $\hat d(x_1,x_2)$ between
two points of space is
\begin{equation}\label{2.6}
\hat d\left(x_1,x_2\right)=\int_{x_1}^{x_2}d\hat s,
\end{equation}
then $\mathcal{B}$ is the border of the metric completion of the domain of
staticity, and this border is unquestionably a 2-dimensional sphere. No
coordinate condition can change that. On the other hand there is no
reason whatsoever to take for granted that~(\ref{2.6}) defines a physical
distance between two points of space, instead of, say, an optical
distance. And if the distance is changed to a new one $\bar d(x_1,x_2)$
which makes of $\mathcal{B}$ a point then the problem of choosing an
appropriate radial coordinate has to be addressed concomitantly with the
new space structure.

We defend indeed in this paper the point of view that~(\ref{2.6}) is an
optical distance and that the physical distance between two points of space
in the static frame of reference is instead
\begin{equation}\label{2.7}
\bar d\left(x_1,x_2\right)=\int_{x_1}^{x_2}d\bar s,
\end{equation}
where $d\bar s$ is the line-element of the principal transform of
(\ref{2.6}). This leads of course to the conclusion that $\Phi$ and
$\Psi$ in~(\ref{1.21}) are respectively the radial and tangential
principal velocities of the speed of light.\footnote{A simplified
and incomplete version of this point of view was developed in
\cite{Bel69} and~\cite{Bel71}.}

To find the principal transform of the exterior Schwarzschild 
solution we shall start with its line-element written in curvature
coordinates~(\ref{2.3}). The general solution of Eq.~(\ref{1.19}) is then~\cite{AT,Pierre}
\begin{equation}\label{2.8}
R(r)=Q_1f_1(r)+Q_2f_2(r),
\end{equation}
$Q_1$ and $Q_2$ being two arbitrary constants of integration and
\begin{equation}\label{2.8.1}
f_1(r)=r-\frac{3m}{2}, \qquad f_2(r)=\sqrt{1-\frac{2m}{r}}\left(r-\frac{m}{2}\right).
\end{equation}
Requiring
\begin{equation}\label{2.9}
\lim_{r\to\infty}R'=1
\end{equation}
leads to
\begin{equation}\label{2.10}
Q_2=1-Q_1.
\end{equation}
As we shall see in Sect.~\ref{sec:matching}, the remaining constant $Q_1$ will be fixed by
matching the exterior Schwarzschild solution to an interior one.
Notice however, as it was already pointed out in~\cite{ABMMR}, the
remarkable fact that if $Q_1=0$ then the domain of staticity of
the exterior Schwarzschild solution is in quo-harmonic coordinates the
interval $R\in\,]0,\infty[$ as for Schwarzschild's historical form. On
the other hand quo-harmonic coordinates are intimately related to
principal transformations and there is no doubt that in the sense of
the l-h-s of~(\ref{1.20}) $R=0$ corresponds intrinsically to a point.

The fate of the status of the Schwarzschild singularity remains
therefore suspended to whatever we can learn from matching the exterior
solution to an interior one using a global system of quo-harmonic
coordinates.


\section{A new perfect fluid, spherically symmetric, static model}
\label{sec:model}

We consider now the field equations~(\ref{1.23}) under the general
assumptions of staticity and spherical symmetry which led to the
line-element~(\ref{1.1}), to which we shall add as a simplifying
assumption the constancy of the energy density:
\begin{equation}\label{3.1}
\rho=\mbox{constant}.
\end{equation}
Using our notations and curvature coordinates Eqs.~(\ref{1.23}) can be written
as the following system of three first-order differential equations:
\begin{equation}\label{3.2}
2B^{-3}B' + r^{-1}\left(1-B^{-2}\right)=\kappa r\rho,
\end{equation}
\begin{equation}\label{3.3}
2B^{-3}(\sigma B)'=\kappa r,
\end{equation}
\begin{equation}\label{3.4}
2AA'+2B^{-1}B' A^2=\kappa r\sigma^{-1}A^2B^2,
\end{equation}
where
\begin{equation}\label{3.5}
\sigma=\frac{1}{p+\rho}.
\end{equation}

Before writing down solutions of these equations let us
remind that we want to obtain solutions that can be matched to
the exterior Schwarzschild solution. This demands the existence of a
value of $r$, say $r_1$, such that

i) on this 2-sphere, say $\mathcal{R}$, the pressure vanishes,
\begin{equation}\label{3.6}
p_1=0,
\end{equation}
ii) and the functions $A$, $B$ are continuous across $\mathcal{R}$.
From~(\ref{2.3}) we have to require then, with obvious notations,
\begin{equation}\label{3.7}
A^{-}_1=A^{+}_1=\sqrt{1-\frac{2m}{r_1}}, \qquad
A^{-}_1B^{-}_1=A^{+}_1B^{+}_1=1.
\end{equation}
We shall discuss the continuity of the derivatives latter on.

The general solution of Eq.~(\ref{3.2}) is
\begin{equation}\label{3.8}
B^{-2}=1-\frac{\kappa}{r}\int_{r_0}^rr^2\rho\,dr
=1-\frac{\kappa\rho}{3r}\left(r^3-r_0^3\right), \qquad r\ge r_0,
\end{equation}
where $r_0$ is an arbitrary constant that we will assume non-negative. 
We shall refer from now on to $r_0$ as the center of symmetry of the 
configuration. That $r_0$ can be understood as being a point and
therefore as a center will be fully justified in Sect.~\ref{sec:matching}.

One has to put $r_0=0$
to obtain the interior Schwarzschild
solution. Since this particular solution satisfies
the regularity conditions
\begin{equation}\label{3.9}
B_0=1, \qquad B'_0=0,
\end{equation}
it has become customary to require the same regularity conditions from
any other physically meaningful solution. This is unjustified for
several reasons.

i) It is assumed implicitly that the range of the radial variable must
be $r\in [0,\infty[$.\footnote{This assumption and $r_0\not=0$ leads to
the singular solution of Volkoff~\cite{Volkoff} and Wyman~\cite{Wyman}.}

ii) An intuitive meaning of $r$ is accepted before knowing whether a
global $\mathcal{C}^1$ model can be completed using curvature coordinates. We
shall see that it can not.

iii) More general regularity conditions can be accepted because they do
not contradict any basic mathematical or physical principle.

iv) It ignores that other interpretations for the
line-element~(\ref{1.2}) are available that do not require these
regularity conditions.

The values of $B$ and its derivative at the center $r=r_0$ are now
\begin{equation}\label{3.11}
B_0=1, \qquad B'_0=\frac12\kappa\rho r_0.
\end{equation}
A more general statement for the derivative $B'$ is that
\begin{equation}\label{3.11.1}
B'=\frac{\kappa\rho}{6r^2}B^3\left(2r^3+r_0^3\right).
\end{equation}
proving that $B$ is a monotonously increasing function.

From~(\ref{3.7}) and~(\ref{3.8}) we obtain the mass or the density depending
on which parameter we may wish to consider as given:
\begin{equation}\label{3.12}
m=\frac16\kappa\rho\left(r_1^3-r_0^3\right), \qquad
\rho=\frac{6m}{\kappa\left(r_1^3-r_0^3\right)}.
\end{equation}

Taking into account~(\ref{3.6}) the solution of~(\ref{3.3}) is
\begin{equation}\label{3.13}
\sigma B=\frac{1}{\rho}B_1-\frac12\kappa\int_r^{r_1}rB^3\,dr.
\end{equation}
The values of $\sigma$ and $\sigma'$ at the center are
\begin{equation}\label{3.13.1}
\sigma_0=\frac{1}{\rho}B_1-\frac12\kappa\int^{r_1}_{r_0}rB^3\,dr, \qquad
\sigma^\prime_0=\frac12\kappa r_0\left(1-\sigma_0\rho\right).
\end{equation}

Taking again into account~(\ref{3.7}) the solution of~(\ref{3.4}) is
\begin{equation}\label{3.14}
A^2=B^{-2}\exp\left(-\kappa\int_r^{r_1}rB^2\sigma^{-1} \,dr\right).
\end{equation}
The values of $A$ and $A'$ at the center are
\begin{equation}\label{3.14.1}
A_0=\exp\left(-\frac12\kappa\int_{r_0}^{r_1}rB^2\sigma^{-1} \,dr\right),
\qquad A^\prime_0=\frac12\kappa r_0 p_0 A_0.
\end{equation}
Notice that neither $\sigma'$, nor $A'$ are zero at the
origin $r=r_0$. This is again a mild departure from conventional
requirements of regularity that does not contradict any mathematical or
physical principle we are aware of.

Let us prove that the pressure $p$ is a monotonously decreasing function
of~$r$. In fact from (\ref{3.13}) and (\ref{3.11.1}), using (\ref{3.5})--(\ref{3.6}) 
it follows that at $r_1$ we have $\sigma_1=\rho^{-1}$ and
\begin{equation}\label{3.14.2}
\sigma^\prime_1=\frac{\kappa}{6r_1^2}B_1^2\left(r_1^3-r_0^3\right) > 0,
\end{equation}
because $r_1> r_0$. Since $\sigma$ is increasing at $r_1$,
it will be monotonously increasing along the whole interval $\left[r_0,r_1\right]$ unless there
is a point $r_0\le r_2<r_1$ where $\sigma'_2=0$ and $\sigma_2 < \sigma_1$, 
which is impossible because if  $\sigma'_2=0$ from (\ref{3.13}) and
(\ref{3.11.1})
it follows that
\begin{equation}\label{3.14.2a}
\sigma_2=\frac{3r_2^3}{2r_2^3+r_0^3}\,\sigma_1\ge\sigma_1.
\end{equation}
This completes 
the proof of our statement. We shall see in Sect. 5 that $p_0$ (and, thus,
the pressure at any other point)  is finite except 
in the most extreme configurations one can think of. We shall 
also discuss there when the condition $p<\rho$ is satisfied.

The behaviour of $A$ and $\sigma$ are very simply related. In fact,
from (\ref{3.13}) and (\ref{3.14}) its follows that $A/\sigma=C$
with $C=\rho\sqrt{1-2m/r_1}$, the value of this constant
being derived from (\ref{3.6}) and (\ref{3.7}).

Let us write $r_0$ in units of $2m$:
\begin{equation}\label{a3.1}
r_0=2m\,a, \qquad a\ge 0.
\end{equation}
Then from (\ref{3.12}) it follows that the mass $m>0$ is the solution of 
the equation
\begin{equation}\label{a3.2}
6m+\kappa\rho a^3(2m)^3=\kappa\rho r_1^3.
\end{equation}
If $0\leq a<1$ then, for any given density, the viable values of $r_1>r_0$ 
are bounded from above. This is a well-known result when $a=0$, 
corresponding to the interior Schwarzschild solution, that follows immediately 
from (\ref{3.8}) but it holds true also when $r_0$ belongs to this more 
general interval. In fact, if we assume that $r_1>2m$, as it is necessary 
from (\ref{3.7}) to guarantee that $B^{-2}_1>0$, then from (\ref{a3.2}) it follows 
that
\begin{equation}\label{a3.3}
\frac{3}{\kappa\rho}>(1-a^3)r_1^2,
\end{equation}
which proves our assertion.

On the contrary, if $a\ge 1$ all values of $r_1$ are viable whatever the 
density. In fact, the values of $m>0$ derived from (\ref{a3.2}) are such 
that $2m <r_1$ because otherwise we would have
\begin{equation}\label{a3.4}
6m<\kappa\rho(1-a^3)(2m)^3\le0,
\end{equation}
in contradiction with the positivity of $m$ derived from (\ref{a3.2}).

The preceding remarks prove that $r_0\>=2m$ selects a distinguished class of 
solutions that allow the consideration of models with arbitrary density, or mass, 
with arbitrary large values of $r_1$.

Consider the sequences of models with different values of $a\ge 1$ that 
can be constructed with a given mass $m$ and decreasing values of $r_1$. 
The sequence
\begin{equation}\label{a3.5}
r_0=2m
\end{equation}
is the only one that allows to reach the limit where $r$ at the exterior covers 
the interval $]2m,\infty[$ of the maximum domain of staticity of the exterior Schwarzschild 
 solution. We shall see in the following sections that this corresponds 
to the most extreme models with point-like sources. And therefore this
is the sequence that allows to construct models with arbitrary mass, or
density, with arbitrary large or small size.

Another striking difference between the two cases $a=0$ and $a >0$ 
comes from the comparison of the geometry
described by the 3-dimensional metric~(\ref{1.2}). Using the orthonormal 
co-basis
\begin{equation}\label{3.14.3.1}
\theta^1=dr, \qquad \theta^2=r\,d\theta \qquad \theta^3=r\sin\theta\, d\varphi,
\end{equation}
the non-zero
components of the Ricci tensor are
\begin{equation}\label{3.14.4}
R_{11}=\frac{\kappa\rho}{3r^3}(2r^3+r_0^3), \qquad 
R_{22}=R_{33}=\frac{\kappa\rho}{6r^3}(4r^3-r_0^3).
\end{equation}
It follows from these formulas the well-known result that for $r_0=0$  
the three non-zero components are identical, meaning that the metric 
(\ref{1.2}) has constant curvature. This is not the case if $r_0>0$ 
whatever this value might be. This means that our solution is 
essentially different from the Schwarzschild one.
 
We postpone the discussion of the physical acceptability of
these solutions as well as the discussion about the geometrical nature
of their center.


\section{Matching interior solutions to the exterior one}
\label{sec:matching}

We have already implemented the continuity of the functions $A$ and $B$
on $r=r_1$. It so happens that there remains no freedom to add new
conditions on the derivatives when using curvature coordinates both in the
interior and the exterior solutions. 
In fact, the derivative of $B$ is
inescapably discontinuous, and the derivative of $A$ is already
continuous without the necessity of requiring it. This is of course the
same that happens when trying to match the exterior and
interior Schwarzschild solutions but let us remind how this comes about.

We shall use again, when necessary, super-indexes $\pm$
to refer to quantities that belong to the exterior or interior
solutions, and the sub-index $1$ to indicate that the corresponding
quantity has been evaluated on the border of the object.
From~(\ref{3.2}) and~(\ref{3.7}) we have
\begin{equation}\label{4.1}
2B_1^{-3}B_1^{-\prime}+\frac{2m}{r_1^2}=\kappa\rho r_1,
\end{equation}
and from~(\ref{2.3}) we have
\begin{equation}\label{4.2}
2B_1^{-3}B_1^{+\prime}+\frac{2m}{r_1^2}=0,
\end{equation}
and therefore inescapably
\begin{equation}\label{4.3}
2B_1^{-3}\left(B_1^{+\prime}-B_1^{-\prime}\right)=-\kappa\rho r_1\not=0.
\end{equation}
Also from~(\ref{3.4})-(\ref{3.7}) it follows that
\begin{equation}\label{4.4}
2A_1A_1^{-\prime}+2B_1^{-3}B_1^{-\prime}= \kappa\rho r_1,
\end{equation}
while from~(\ref{2.3}) we have
\begin{equation}\label{4.5}
2A_1A_1^{+\prime}+\frac{2m}{r_1^2}=0.
\end{equation}
Subtracting~(\ref{4.4}) from~(\ref{4.1}) and using the last result proves as
stated that
\begin{equation}\label{4.6}
2A_1\left(A_1^{+\prime}-A_1^{-\prime}\right)=0.
\end{equation}

Because of~(\ref{4.3}) the model thus constructed is not of class 
$\mathcal{C}^1$ across the border of the object. Most authors of papers and books
when dealing with Schwarzschild's solutions feel satisfied with an
identical situation. This is a mistake because it perverts the nice
axiomatization of General Relativity proposed by Lichnerowicz
\cite{Lichnerowicz} and introduces an ambiguity in the theory if one
wants to have an interpretation for the metric~(\ref{1.2}), and the
coefficients~(\ref{1.21}), as we have for instance for the function $A$.
Besides,  it is well known\footnote{See for example
\cite{Misner} or~\cite{Bel67}.} that when~(\ref{3.6}) holds then an interior
static spherical solution can always be matched to the exterior
Schwarzschild metric. Below we prove it for our particular interior
solution.\footnote{Another example can be seen in~\cite{Liu}.}

Let us consider the differential equation~(\ref{1.19}) with
$B=B^-$ given by Eq.~(\ref{3.8}). The expansion of $B$ around $r=r_0$ is
\begin{equation}\label{4.8.1}
B=1+\frac12\kappa\rho r_0\left(r-r_0\right)+O\left[(r-r_0)^2\right],
\end{equation}
and therefore we have to distinguish two cases.

i) If $r_0=0$, i.e., when the metric is the interior Schwarzschild solution 
then $r=r_0=0$ is a regular singular point and the solutions that are
finite 
at this origin are all proportional to the following one:\footnote{This case 
has been considered 
by P. Teyssandier~\cite{Pierre}.}
\begin{equation}\label{4.8.1a}
R=\frac{3}{2qr^2}\left[\arcsin(\sqrt{q}r)-\sqrt{q} r\sqrt{1-qr^2}\right],
\qquad q\equiv\frac13\kappa\rho.
\end{equation}
These finite solutions are such that
\begin{equation}\label{4.8.2}
R^-(0)=R^-_0=0, \qquad R^{-\prime}(0)=R^{-\prime}_0,
\end{equation}
the derivative at the origin $R^{-\prime}_0$ remaining arbitrary at
this point.

ii) If $r=r_0>0$ then the origin is an ordinary
point of the differential equation. 
Therefore we can choose
arbitrarily the values of $R$ and its derivative at this point:
\begin{equation}\label{4.7}
R^-\left(r_0\right)=R^-_0, \qquad R^{-\prime}\left(r_0\right)=R^{-\prime}_0.
\end{equation}
Each solution $R(r)$ of
the differential equation defines new coordinates~(\ref{1.18}) which are
quo-harmonic for the interior solution. But we do not have, and we
shall not use, their explicit analytic expression.

These initial conditions can be chosen such that if $R^+$ is the
solution  of the same equation~(\ref{1.19}), with $B=B^+$ given by Eq.
(\ref{2.3}), corresponding to initial conditions on the border $r=r_1$,
\begin{equation}\label{4.8}
R^+_1=R^-_1 \qquad R^{+\prime}_1=R^{-\prime}_1,
\end{equation}
then the global model defined by the coordinate
transformation $R=R^+(r)$ at the
exterior and $R=R^-(r)$ in the interior is of class $\mathcal{C}^1$ across
the border. In fact let the new expression of the metric~(\ref{1.2}) be
\begin{equation}\label{4.10}
d\hat s^2=\tilde B^2dR^2+\tilde D^2 R^2d\Omega^2,
\end{equation}
where
\begin{equation}\label{4.11}
\tilde B=\frac{B}{R'}, \qquad \tilde D=\frac{r}{R}.
\end{equation}
$\tilde B$,  $\tilde D$ and $\tilde D'$ are continuous because $R$
and $R'$ have been chosen to be continuous. From the equation
(\ref{1.19}) that satisfy both $R^\pm$ we can derive that
\begin{equation}\label{4.12}
R_1^{+\prime\prime}-R_1^{-\prime\prime}
-2B_1^{-1}\left(B_1^{+\prime}-B_1^{-\prime}\right)R_1'=0.
\end{equation}
From~(\ref{4.11}) we have
\begin{equation}\label{4.14}
\tilde B'=B' R'^{-1}
-BR'^{-2}R''.
\end{equation}

Therefore,
\begin{equation}\label{4.15}
\frac{d\tilde B}{dR}=\tilde B' R'^{-1}=
R'^{-2}B'-R'^{-3}B R''
\end{equation}
and
\begin{equation}\label{4.13}
\frac{d\tilde B^+_1}{dR}-\frac{d\tilde B^-_1}{dR}=
R'^{-2}_1\left(B^{+\prime}_1-B^{-\prime}_1\right)-
R_1'^{-3}B_1 \left(R_1^{+\prime\prime}-R_1^{-\prime\prime}\right).
\end{equation}
Using~(\ref{4.12}) in this equation we finally obtain
\begin{equation}\label{4.16}
\frac{d\tilde B^+_1}{dR}-\frac{d\tilde B^-_1}{dR}=0,
\end{equation}
which proves that the global quo-harmonic model is of class $\mathcal{C}^1$.

Once $R_1^-$ and $R_1^{-\prime}$ have been found integrating~(\ref{1.19}) with
initial conditions~(\ref{4.7}) the constants $Q_1$ and $Q_2$ of~(\ref{2.8})
can be found solving the system of linear equations
\begin{equation}\label{4.17}
R_1^- = Q_1f_1(r_1)+Q_2f_2(r_1), \qquad
R_1^{-\prime} = Q_1f'_1(r_1)+Q_2f'_2(r_1).
\end{equation}
Since the two functions $f_1$ and $f_2$ are independent this system
has always a solution. 

In case i) above the supplementary condition~(\ref{2.10}) determines the
remaining arbitrary constant $R_0^{-\prime}$, while in case ii) this
condition gives only a relation between the two initial
conditions~(\ref{4.7}). To obtain a second relation and make our model
completely determine let us consider the principal transform of the interior metric of
space. By definition we shall have~(\ref{1.20}) with $R=R^-$ and
appropriate functions $\Phi$, $\Psi$, $B^-$ and $C^-$. If we accept,
as we have done, that the structure of space in the static frame of
reference of the source is given by the l-h-s of this equation and we
want to have a center for our configuration and no hole in the space
we are forced to choose our function $R$ with the initial condition
\begin{equation}\label{4.18}
R^-\left(r_0\right)=R^-_0=0.
\end{equation}

Since the functions $R(r)$ and $r(R)$ are not explicitly known when $r_0>0$, 
neither is known explicitly the space-time metric (\ref{1.1}) using
the radial coordinate~$R$. Therefore it is not possible using $R$ to
discuss directly the behavior of the causal geodesics that go through the
center of the configuration. But this can be done looking first at this
problem using the original curvature coordinate $r$. Let us consider a
causal geodesic reaching $P_\mathrm{in}$ with $r=r_0$ at some time $\tilde{t}$. 
Since the space trajectory is contained in a plane, say $\Pi$, of the
auxiliary Euclidean space with line-element $dr^2+r^2d\Omega^2$, we can
use polar coordinates such that $\Pi$ is the plane $\theta=\pi/2$. Let
$\tilde{\varphi}$ be the value of the azimuth angle at $P_\mathrm{in}$; 
and let $\dot r=\tilde{v}_r$ and $\dot\varphi=\tilde{v}_\varphi$ be the 
two corresponding
derivatives. Let us consider now the causal geodesic starting at time $\tilde{t}$
 from the point $P_\mathrm{out}$ defined by
$r=r_0$ and $\varphi=\varphi+\pi$ with derivatives
given by $\dot r=-\tilde v_r$ and $\dot\varphi=-\tilde v_\varphi$. 
The space-time trajectory of this geodesic branch is contained also on $\Pi$
and is the symmetric image of the incident branch with respect to the
diameter joining $P_\mathrm{in}$ and $P_\mathrm{out}$. We see therefore that the
process of contraction of the sphere $r=r_0$ to the point $R=0$
described by (\ref{1.18}) joins the end-point of the incoming branch of
the geodesic with the origin of the outgoing one. And in the process the
matched geodesic is as smooth as they were the two branches. Notice also
that, since starting from any point there are always a bunch of geodesics
reaching $r=r_0$ with different values of $\theta$ and $\varphi$, our construction of the center of the new models endows this center
with a focalising property.

The quo-harmonic class of coordinates derived from the radial coordinate $R$ are natural
coordinates in the sense that they have the following properties: i)
they exhibit the center as a point in the sense of (\ref{1.13}) and ii)
they make the space-time metric smooth across the border of the spherical
source. Another system of natural coordinates sharing these two
properties are the Gauss coordinates derived from the radial coordinate:
\begin{equation}\label{4.19}
R_G=\int_{r_0}^r B\,dr.
\end{equation}
Instead for large values of $r$ the quo-harmonic $R$ behaves as
\begin{equation}\label{4.20}
R=r-\frac32m+O(1/r^2),
\end{equation}
while the Gaussian $R_G$ behaves as
\begin{equation}\label{4.21}
R=r-m\left[1+\ln\left(\frac{m}{2r}\right)\right]+O(1/r),
\end{equation}
which makes the asymptotic behaviour of the metric to depart a little bit more 
from the Newtonian intuition.

In the next section we shall discuss the numerical
solutions and in the final section we shall discuss the physical
relevance of them.


\section{Numerical analysis of the model}
\label{sec:numer}

The numerical study of the model is greatly simplified by using
the following dimensionless variables:
\begin{equation}\label{eq:defx}
x\equiv\frac{r}{2m},\qquad a\equiv x_0=\frac{r_0}{2m},\qquad b\equiv x_1=\frac{r_1}{2m}.
\end{equation}
Now, using~(\ref{3.12}) the expression~(\ref{3.8}) reduces to
\begin{equation}\label{eq:Bx}
B^{-2}=1-\frac1x\,\frac{x^3-a^3}{b^3-a^3}.
\end{equation}
%

\subsection{The pressure}

For the interior solution corresponding to $x\in[a,b]$, 
Eq.~(\ref{3.13}) reduces to
\begin{equation}\label{eq:mu}
\mu\equiv\frac{\sigma}{4\kappa m^2}=B^{-1}\left[\frac13\sqrt{\frac{b}{b-1}}\left(b^3-a^3\right)-\frac12
\int_x^bxB^3\,dx\right],
\end{equation}
which can be written in terms of elliptic functions, but is far more easily 
computed numerically. 

\begin{figure}[ht]\centering
\includegraphics[width=\textwidth]{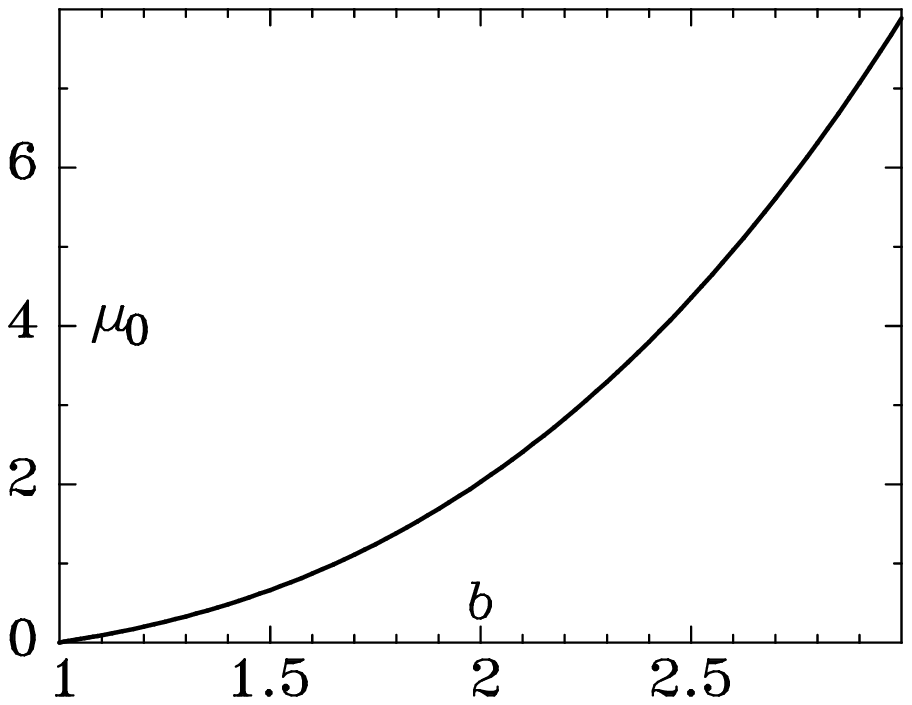}
\caption{Function $\mu_0$ for $a=1$ and small values of $b$.\label{figmu0}}
\end{figure}

\begin{figure}[ht]\centering
\includegraphics[width=\textwidth]{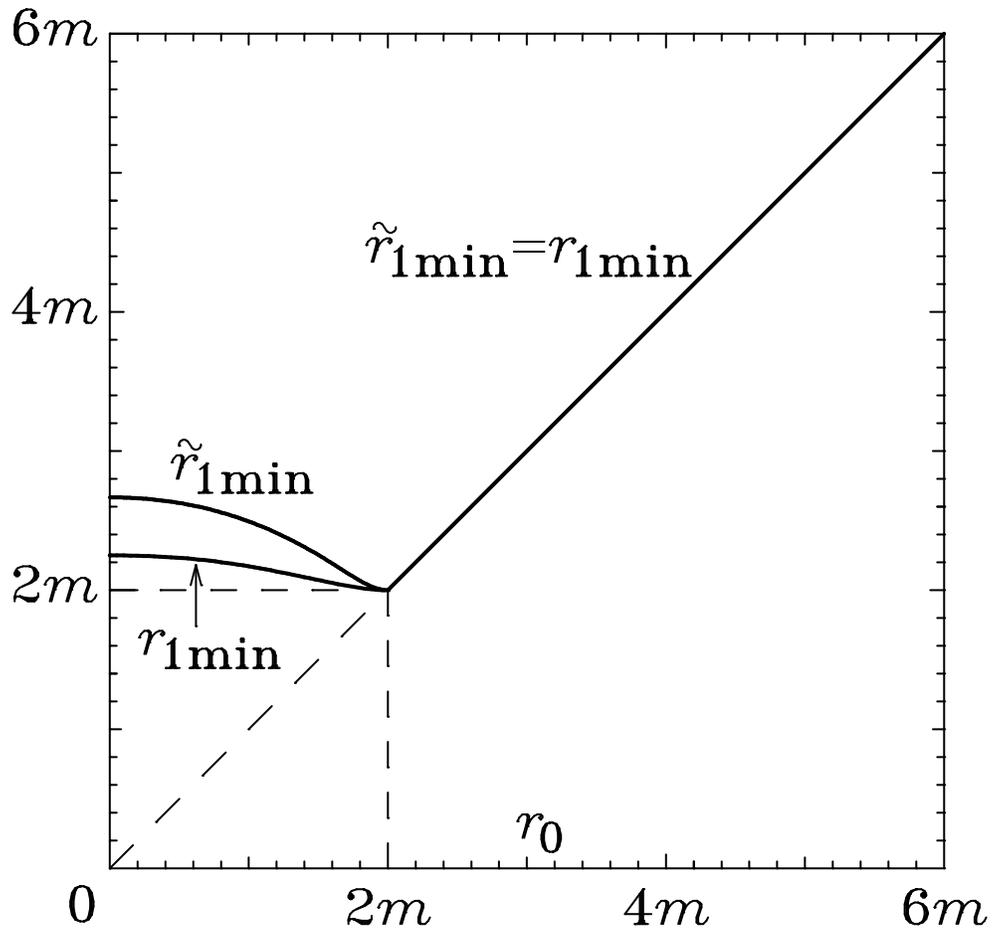}
\caption{Values of $r_1$ for which the pressure
becomes infinite ($r_{1\mathrm{min}}$) or equal to $\rho$ ($\tilde{r}_{1\mathrm{min}}$) 
at the origin~$r_0$.\label{figr0min}}
\end{figure}

Since the pressure decreases from the
center $r=r_0$ to the border $r=r_1$, to make sure that it does not go to infinity
it is enough to check that it does not diverge at $r=r_0$, i.e., that
$\mu_0\equiv\mu(a)$ does not become zero (or negative).
This  can be easily checked by numerical quadrature. In Fig.~\ref{figmu0}
we see that, in the case $a=1$ ($r_0=2m$),  $\mu_0$ only vanishes for $b=a$,
which corresponds to the limit case in which the interior reduces to a point. The
same happens for $a>1$, but for $0\le a<1$ the pressure becomes infinite
at some interior point 
unless the matching radius $r_1=2mb$ is greater than a given value $r_{1\mathrm{min}}$, 
which is displayed in Fig.~\ref{figr0min}. In the case $r_0=0$ corresponding to 
 the interior Schwarzschild solution, the well-known minimum value is 
$r_{1\mathrm{min}}=\frac94m$.

One may also check numerically that the dominant energy condition 
$p<\rho$ ---i.e., $\mu\ge\mu_0>\frac12\mu_1\equiv\frac12\mu(b)$--- is satisfied always
if $r_0\ge 2m$ and $r_1>r_0$. For $0\le r_0<2m$, however, it is satisfied
only for values of $r_1$ greater than a minimum value $\tilde{r}_{1\mathrm{min}}$ which
happens to be higher than the $r_{1\mathrm{min}}$ discussed above, as
displayed in Fig.~\ref{figr0min}. In the case  of
 the interior Schwarzschild solution ($r_0=0$) one has 
$\tilde{r}_{1\mathrm{min}}=\frac83m$.

Thus, we conclude that $2m$ is the smallest value of $r_0$ for which the physical
conditions on the pressure do not put any limit on the matching radius $r_1>r_0$.

\subsection{Quo-harmonic coordinates}

To compute the quo-harmonic coordinates~(\ref{1.18}) we have to
solve the differential equation~(\ref{1.19}), which in the dimensionless
variables of this section reduces to
\begin{equation}\label{eq:difeq}
2xHS''+\left(4H-2x^3-a^3\right)S'-4\left(b^3-a^3\right)S=0,
\end{equation}
where 
\begin{equation}\label{eq:Sdef}
S\equiv\frac{R}{2m},
\end{equation}
a prime indicates derivative with respect to $x$ and we have defined
\begin{equation}\label{eq:defH}
H\equiv  x \left(b^3-a^3\right)B^{-2}=x \left(b^3-a^3\right)-\left(x^3-a^3\right)>0.
\end{equation}

In the case $a=0$ (corresponding to 
 the interior Schwarzschild solution), the origin is regular singular, but the solution
satisfying $\tilde{S}(a)=0$, $\tilde{S}'(a)=1$
exists in $[0,b]$ and can be explicitly written~\cite{Pierre} as~(\ref{4.8.1a}), which reduces to
\begin{equation}
\tilde{S}=\frac{3b^3}{2x^2}\left[b^{3/2}\arcsin\left(b^{-3/2}x\right)-x\sqrt{1-b^{-3}x^2}\right].
\end{equation}

In the remaining cases ($a>0$), the differential equation is linear, the coefficients are
continuous, and $H$ does not vanish for $x\in[a,b]$ (we are excluding the limit case 
$b=a$), so that there exists a unique solution defined for $a\le x\le b$ that satisfies
the initial conditions $S(a)=0$, $S'(a)=\beta$,
where $\beta$ is a constant to be computed later. In fact, $S=\beta \tilde{S}$, if
$\tilde{S}$ is the solution of Eq.~(\ref{eq:difeq}) satisfying
\begin{equation}\label{eq:initcondS}
\tilde{S}(a)=0,\qquad \tilde{S}'(a)=1.
\end{equation}

\begin{figure}[ht]\centering
\includegraphics[width=\textwidth]{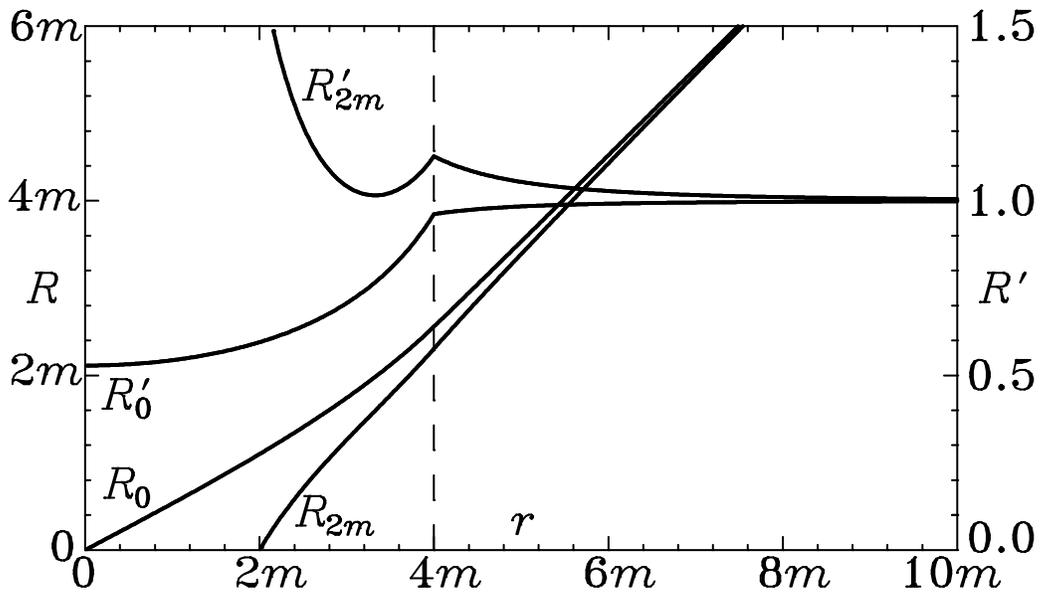}
\caption{Function $R$ and its derivative $R'$ for $r_0=0,2m$ and $r_1=4m$.\label{figssd}}
\end{figure}

Now the numerical method to compute $R$ is straightforward. After selecting the parameter $a$,
we compute $\tilde{S}$ by solving~(\ref{eq:difeq}) with~(\ref{eq:initcondS}). Then
the auxiliary condition~(\ref{2.10}) and the matching conditions~(\ref{4.17})
read
\begin{equation}\label{eq:matc}
\beta\tilde{S}(b) = Q_1g_1(b)+\left(1-Q_1\right)g_2(b), \qquad
\beta\tilde{S}'(b) = Q_1g_1'(b)+\left(1-Q_1\right)g_2'(b),
\end{equation}
where functions~(\ref{2.8.1}) are written as
\begin{equation}\label{eq:f12a}
g_1(x)\equiv\frac{f_1}{2m}=x-\frac{3}{4}, 
\qquad g_2(x)\equiv\frac{f_2}{2m}=\sqrt{1-\frac{1}{x}}\left(x-\frac{1}{4}\right).
\end{equation}
Since Eqs.~(\ref{eq:matc}) are readily solved for $\beta$ and $Q_1$, $R$ is given by
\begin{equation}
S=\frac{R}{2m}=\left\{
\begin{array}{ll}
\beta \tilde{S}(x),&\mbox{for }a\le x \le b,\\
Q_1g_1(x)+\left(1-Q_1\right)g_2(x),&\mbox{for } x \ge b.
\end{array}
\right.
\end{equation}
Two particular cases, for $a=0,1$ and $b=2$ (i.e., for $r_0=0,2m$ and $r_1=4m$) 
are displayed in Fig.~\ref{figssd}. Notice the different definition ranges of $R$ in
curvature coordinates.

\begin{figure}[ht]\centering
\includegraphics[width=\textwidth]{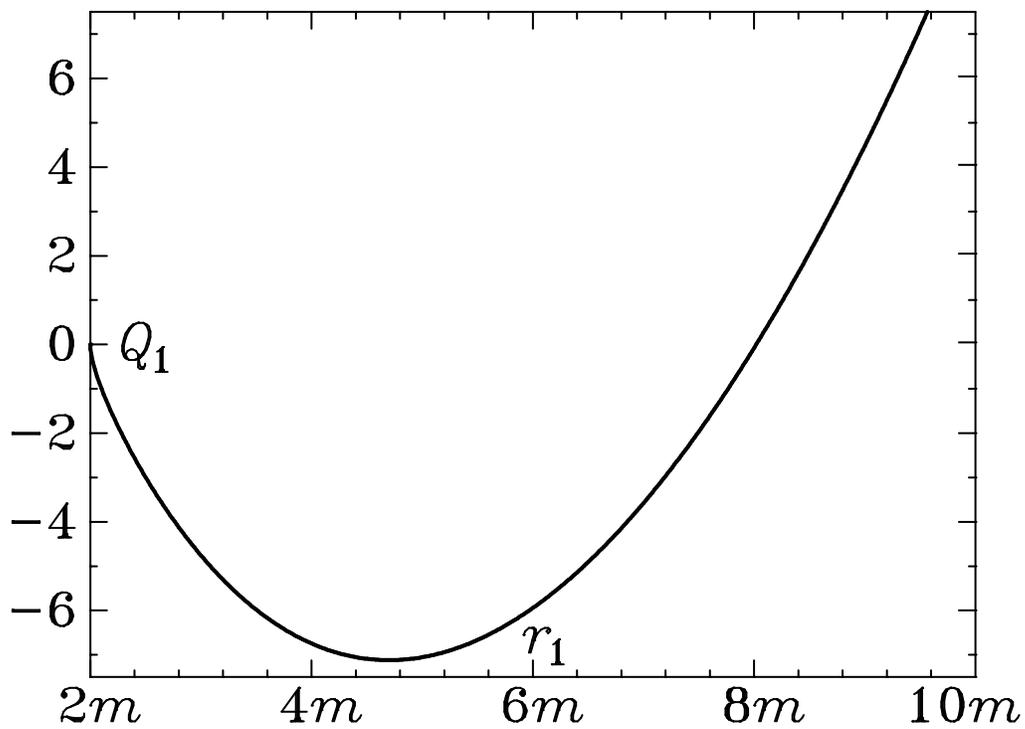}
\caption{Coefficient $Q_1$ for $r_0=2m$ and different values of $r_1$.\label{figq1}}
\end{figure}

In Fig.~\ref{figq1} the coefficient $Q_1$ is displayed for $a=1$ and
different values of~$b$. One may see there that in the limit $b\to a=1$, i.e., when the interior
collapses to the point $R=0$ in quo-harmonic coordinates and to the center $r_0=2m$ in curvature
coordinates, $Q_1$ vanishes, so that the radial coordinate $R$ associated to
quo-harmonic coordinates is given everywhere by $R=f_2(r)$,
as pointed in Sect.~\ref{sec:model}.

\subsection{Metric coefficients and principal transform}

The metric coefficient $B$ is given, in the dimensionless curvature coordinates, by~(\ref{eq:Bx})
for $a\le x\le b$ and by $B=\left(1-1/x\right)^{-1/2}$ for $x\ge b$. Similarly, 
the other metric component in curvature coordinates is given by
Eq.~(\ref{3.14}), which reduces to
\begin{equation}\label{eq:a2bm2}
A^2=B^{-2}\exp\left(-\int_x^{b}xB^2\mu^{-1} \,dx\right),
\end{equation}
for $a\le x\le b$ and is $A=B^{-1}$ for the exterior $x\ge b$.
Since we can compute numerically the quo-harmonic coordinates, they can be used to display
the metric of different models in the same physical coordinates. 

The functions in Eqs.~(\ref{1.21}) now are written as
\begin{equation}\label{eq:phipsi}
\Phi=\frac{S'}{B}, \qquad \Psi=\frac{S}{x}
\end{equation}
and can be readily computed.

\begin{figure}[ht]\centering
\includegraphics[width=\textwidth]{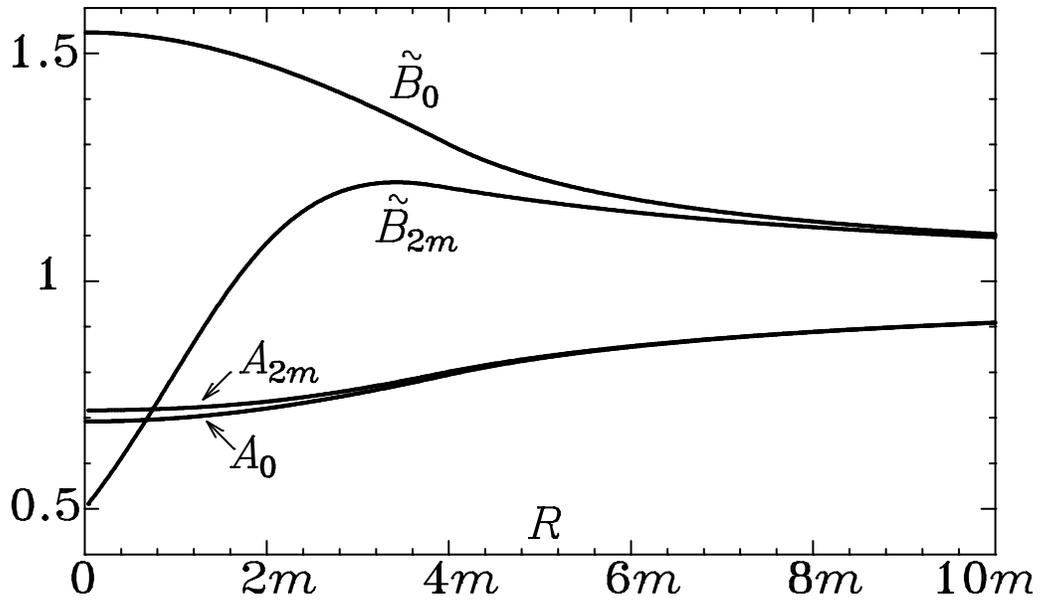}
\caption{Metric coefficients $A$ and $\tilde{B}$ when the
matching radius is $R_1=4m$ and $r_0=0,2m$.
The same $R$ coordinate associated to quo-harmonic coordinates is used in both cases.\label{figab}}
\end{figure}

\begin{figure}[ht]\centering
\includegraphics[width=\textwidth]{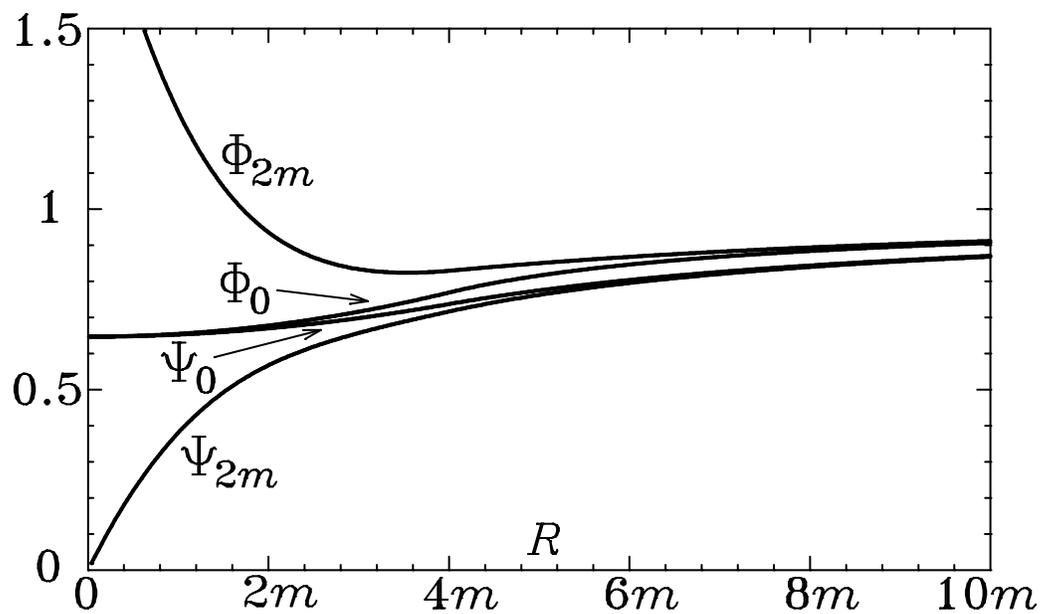}
\caption{Functions $\Phi$ and $\Psi$ when the
matching radius is $R_1=4m$ and $r_0=0,2m$.
The same $R$ coordinate associated to quo-harmonic coordinates is used in both cases.
\label{figphipsi}}
\end{figure}

For instance, in
Fig.~\ref{figab} we show the metric coefficients $A$ and $\tilde{B}$ for two models with $a=0,1$
(i.e., $r_0=0,2m$). In both cases the matching of the interior and exterior
metrics happens at the same location in quo-harmonic coordinates: at the spherical
surface   of radius $S=2$ ($R=4m$), which in curvature coordinates has radius
$x\approx2.7134$ when $a=0$ and $x\approx2.7916$ if $a=1$. The continuity of the metric 
coefficients and their first derivatives is apparent in the figure. The functions
$\Phi$ and $\Psi$ corresponding to the same special cases are plotted in 
Fig.~\ref{figphipsi}.


\section{Concluding remarks}
\label{sec:remarks}

Among all possible values of $r_0$ two of them are distinctly
distinguished:

i) the value $r_0=0$ because it leads to the most well-behaved models,
although only in a restrictive range of the density $\rho$ and  the radius of the interior configuration
$r_1$
or $R_1$, and	

ii) the value $r_0=2m$  because this leads to the closest sequence to
Schwarzschild's one, allows unrestricted values of the parameters $\rho$
and $r_1$ or $R_1$ and contains the emblematic most extreme 
configuration we can think of as a limit, namely that with a point-like
source.

Whether or not new considerations or difficulties will suggest selecting
or excluding particular values of $r_0$ remains to be seen. But if only
one new value remains then be pretty sure that this value will be $r_0=2m$. 

There exists a range of parameters $\rho$ and $r_1$ for which
Schwarzschild's
interior solution and ours coexist. Nevertheless the two solutions are quite
different, whatever
these parameters might be. This is so because the geometry described by
(\ref{1.2}) is homogeneous and isotropic in one case and inhomogeneous
and anisotropic in the other.

We have assumed for simplicity that the energy-density was constant. It
might be objected that this is not a realistic ``equation of state" to
describe the properties of our objects, and in particular the
extreme ones, i.e., those without analog based on the interior Schwarzschild
solution. But no equation of state will be realistic, and would
be un-realistic to guess one, as long as new physics is not available
for them. Not to mention that to assume the existence of an equation of
state is also a simplifying assumption, as it is to use a perfect fluid
description. The main properties of our models do not depend crucially,
we believe, on these simplifying assumptions. They depend instead
crucially on the identification of $r=2m$ with the origin and the
center of of symmetry of the model and also on the re-interpretation of
the metric (\ref{1.2}) based on the consideration of principal
transformations.

We do not claim that such objects will be
found in nature, but we do claim that if they are found General Relativity is a
good theory to understand its geometrical properties and gravitational
field. We also claim that it is a sound attitude to look for them.
Everybody knows where there is a good chance to discover them.

\section*{Acknowledgments}

We acknowledge a careful reading of the first version of our manuscript
by J.~M.~M.~Senovilla and his constructive criticisms which led to a
few improvements included in this second one.

The work of JMA was supported by the University of the Basque Country
through the research project~UPV172.310-EB150/98 and the General Research
Grant~UPV172.310-G02/99. Ll.~Bel gratefully acknowledges
as visiting professor the hospitality of the UPV/EHU.

\newpage


\newpage

\newpage

\newpage

\newpage

\newpage

\vfill

\end{document}